\documentclass[sigconf,screen=true,9pt]{acmart}

\usepackage{enumitem}
\setlist{leftmargin=5.08mm}

\usepackage{lipsum}
\usepackage{graphicx}
\usepackage{amsmath}
\usepackage{footnote}

\usepackage{mathtools}
\usepackage{mathrsfs}     
\usepackage{comment}
\usepackage{soul}                                         
\usepackage[subrefformat=parens,labelformat=parens]{subfig}
\usepackage{bm,bbm}
\usepackage{multirow}
\usepackage{booktabs}
\usepackage{blkarray}
\usepackage{tikz}
\usetikzlibrary{positioning,calc,fit,decorations.pathmorphing,shapes.geometric,shapes.gates.logic.US,calc}
\usetikzlibrary{arrows,arrows.meta,decorations.markings,shapes,shapes.arrows}
\usetikzlibrary{patterns,snakes}
\usetikzlibrary{decorations,decorations.pathreplacing}
\usetikzlibrary{backgrounds}
\usepackage{tikz-3dplot}
\usepackage{balance}

\usepackage{courier}                                       
\usepackage{cleveref}                                   

\usepackage[mathcal]{eucal}

\usepackage[]{algpseudocode}                               

\makeatletter\let\float@addtolists\relax\makeatother
\usepackage{algorithm}

\usepackage{filecontents}                                  
\usepackage{pgfplots}
\usepackage{pgfplotstable}
\usepackage{pgf-pie}
\usepgfplotslibrary{groupplots}
\pgfplotsset{compat=newest}
\usepackage[figuresright]{rotating}
\usepackage{xcolor,colortbl}                             
\usepackage{makecell}
\usepackage{siunitx}
\usepackage{color}
\usepackage{diagbox}

\theoremstyle{plain}

\theoremstyle{definition}

\definecolor{CUHKorange}{RGB}{244,106,18} 
\definecolor{CUHKblue}{RGB}{0,111,190}    
\definecolor{CUHKgreen}{RGB}{0,127,128}   
\definecolor{CUHKred}{RGB}{228,46,36}     
\definecolor{CUHKyellow}{RGB}{198,148,34} 
\definecolor{CUHKdark}{RGB}{114,44,114}   
\definecolor{CUHKmiddle}{RGB}{144,44,144} 

\graphicspath{{./}}

\copyrightyear{2027}
\acmYear{2027}
\setcopyright{cc}
\setcctype{by}
\acmConference[ASPDAC '27]{32nd Asia and South Pacific Design Automation Conference}{January 25--28, 2027}{Tokyo, Japan}
\acmBooktitle{32nd Asia and South Pacific Design Automation Conference (ASPDAC '27), January 25--28, 2027, Tokyo, Japan}
\acmDOI{10.1145/3840404.3848371}
\acmISBN{979-8-4007-2923-2/2027/01}

\begin{document}

\title{VeriGRAG: Enhancing LLM-Based Verilog Code Generation with Structure-Aware Soft Prompts}


\author{Jiayu Zhao}
\orcid{0009-0005-8138-8565}
\affiliation{%
  \institution{University of Science and Technology of China}
  \city{Hefei}
  \country{China}
}
\email{jiayz@mail.ustc.edu.cn}

\author{Zehua Pei}
\orcid{0000-0002-2884-2459}
\affiliation{%
  \institution{The Chinese University of Hong Kong}
  \city{Hong Kong}
  \country{China}
}
\email{zehuapei@link.cuhk.edu.hk}

\author{Zetao Guo}
\orcid{0009-0006-3128-4618}
\affiliation{%
  \institution{University of Science and Technology of China}
  \city{Hefei}
  \country{China}
}
\email{ztguo@mail.ustc.edu.cn}

\author{Bei Yu}
\orcid{0000-0001-6406-4810}
\affiliation{%
  \institution{The Chinese University of Hong Kong}
  \city{Hong Kong}
  \country{China}
}
\email{byu@cse.cuhk.edu.hk}

\author{Song Chen}
\authornote{Corresponding author.}
\orcid{0000-0003-0341-3428}
\affiliation{%
  \institution{University of Science and Technology of China}
  \city{Hefei}
  \country{China}
}
\email{songch@ustc.edu.cn}

\renewcommand{\shortauthors}{Zhao et al.}


\begin{abstract}
Large language models (LLMs) have demonstrated remarkable capabilities in generating Verilog code from natural language descriptions. However, Verilog code inherently encodes the structural characteristics of hardware circuits, including data path organization and interconnections among components. Effectively leveraging these structural characteristics to enhance the functional and syntactic correctness of LLM-based Verilog code generation remains a significant challenge. To address this challenge, we propose VeriGRAG, a novel framework that converts Verilog code into data path graphs and employs graph neural networks (GNNs) to encode their structural information. A multimodal retriever, trained through knowledge distillation from a cross-attention encoder to a dual encoder, then selects the graph embeddings most relevant to the input hardware description. The VeriFormer module uses learnable query tokens and two-stage training to extract code-relevant structural features and align them with the LLM's input embedding space, producing structure-aware soft prompts. Experiments on VerilogEval and RTLLM show improvements in both syntactic and functional correctness. With Qwen2.5-Coder-32B as its backbone, VeriGRAG achieves the highest functional $pass@1$ and $pass@5$ scores among the compared methods on both benchmarks.
\end{abstract}



\maketitle


\section{Introduction}

Digital circuit design is a demanding and labor-intensive process, particularly when translating functional specifications into register-transfer level (RTL) designs. As RTL designs strongly influence subsequent stages of the electronic design automation (EDA) flow, automating Verilog code generation at this stage can substantially improve design efficiency. Recent advances in LLMs, particularly commercial models such as GPT-4o~\cite{hurst2024gpt} and Claude-3~\cite{anthropic2024claude3}, have demonstrated strong code generation capabilities. However, using closed-source services for Verilog generation raises concerns about intellectual property protection and limits domain-specific customization. Open-source code models such as CodeLlama~\cite{roziere2023code}, DeepSeek-Coder~\cite{guo2024deepseek}, and Qwen2.5-Coder~\cite{hui2024qwen2} therefore offer a practical alternative by enabling private deployment and task-specific adaptation.

Despite recent advances in LLM-based Verilog code generation, several critical challenges remain unresolved. Unlike software code, which typically follows a sequential execution paradigm, Verilog is inherently designed to describe hardware circuit structures. It comprises modules representing physical components that frequently execute concurrently. Furthermore, Verilog encompasses a diverse set of constructs, including combinational and sequential logic, finite-state machines, and hierarchical module interconnections, which collectively capture both behavioral and structural aspects of hardware designs. The relative scarcity of high-quality Verilog code corpora for training further exacerbates the difficulty of understanding and generating hardware description code. These challenges are particularly pronounced in the context of Verilog programs exhibiting concurrency and structural complexity. Therefore, advancing the application of LLMs in Verilog code generation necessitates effective modeling of structural information.

\begin{figure}[t]
    \centering
    \includegraphics[width=1.0\columnwidth]{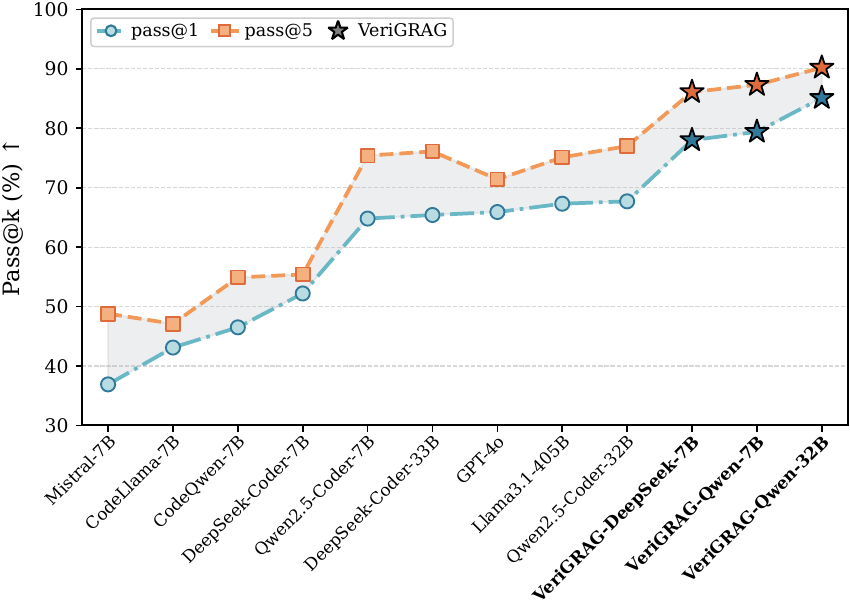}
    \caption{Performance comparison of different base models and VeriGRAG on the VerilogEval-Machine~\cite{liu2023verilogeval} benchmark.}
    \Description{Comparison of pass@1 and pass@5 scores on VerilogEval-Machine. VeriGRAG improves both metrics over the corresponding base models across different LLM backbones.}
    \label{fig_comparison}
\end{figure}

Recent studies have actively explored open-source LLMs for Verilog generation, improving their capabilities through data synthesis, retrieval augmentation, self-reflection, multi-agent collaboration, and functional feedback~\cite{liu2024rtlcoder,pei2024betterv,gao2024autovcoder,cui2024origen,zhao2024mage,mi2024promptv,yang2025haven,thakur2023autochip,ho2025verilogcoder}. These efforts demonstrate the potential of open-source models for Verilog generation.

However, existing approaches do not fully exploit the structural information inherent in hardware designs, including circuit topology, datapath organization, and dependencies among hardware modules. Most methods represent hardware specifications and code as monolithic text, thereby obscuring these structural cues. This motivates the following research question:

\begin{center}
    \emph{How can we leverage the intrinsic structural information in Verilog code to improve LLM-based Verilog generation?}
\end{center}

\begin{figure*}[th!]
    \centering
    \includegraphics[width=0.74\textwidth]{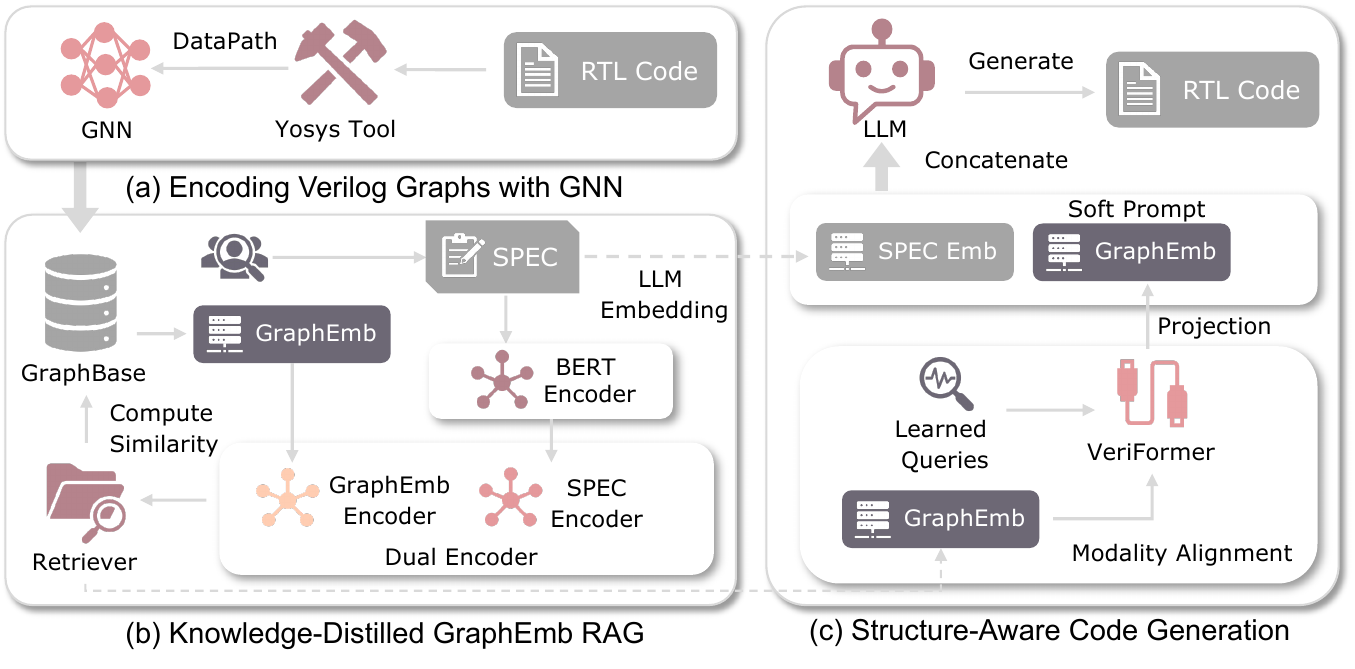}
    \caption{
        Overview of the Structure-Aware VeriGRAG Framework.
        (a) We first extract data path graphs from Verilog code using Yosys and encode them with a graph neural network (GNN).
        (b) Given a user-provided hardware description~(SPEC), we retrieve the most relevant graph embedding from the graph database.
        (c) VeriFormer aligns the graph embedding with the LLM's input embedding space and provides structure-aware soft prompts to enhance the LLM's capability in generating accurate Verilog code.
    }
    \label{fig_overview}
    \Description{Verilog code is converted into data path graphs using Yosys and encoded by a graph neural network. A hardware description is used to retrieve the most relevant graph embedding from the graph database. VeriFormer aligns the retrieved embedding with the LLM's input embedding space to provide structure-aware soft prompts for Verilog code generation.}
\end{figure*}

In this paper, we propose VeriGRAG, a framework that enhances the capability of large language models (LLMs) in generating Verilog code by explicitly incorporating structural information. Extensive experiments demonstrate that VeriGRAG consistently outperforms existing baselines, achieving the best overall performance on both VerilogEval~\cite{liu2023verilogeval} and RTLLM~\cite{lu2024rtllm} benchmarks. 
As shown in Figure~\ref{fig_comparison}, VeriGRAG improves both $pass@1$ and $pass@5$ over the corresponding base models on VerilogEval-Machine, demonstrating its effectiveness across different LLM backbones.
The overall framework is illustrated in Figure~\ref{fig_overview}.
We employ Yosys~\cite{wolf2016yosys} to extract data path graphs from Verilog code, encode them using graph neural networks (GNN), and store the resulting graph embeddings in a graph database for efficient retrieval.
We also train a multimodal retriever to retrieve the most similar graph embeddings from the graph database based on user-provided hardware descriptions.
These graph embeddings are subsequently used to provide structural information to the LLM.
Inspired by~\cite{li2023blip}, we introduce VeriFormer, a module that employs learnable query tokens to extract structure-informative features from graph embeddings and align them with the LLM’s feature space,
providing structure-aware soft prompts to enhance the generation of high-quality Verilog code. 

The contributions of this paper are as follows:
\begin{itemize}
    \item We present a novel approach for extracting structural features from Verilog code, where Yosys is employed to transform the code into data path graphs. These graphs are then encoded into structured graph embeddings using GNN. 
    \item We propose a multimodal retrieval training paradigm that leverages knowledge distillation to train a dual encoder, enabling it to inherit the strong cross-modal alignment capabilities of a cross-attention encoder. This approach preserves the efficiency of the dual encoder architecture by caching graph embeddings for fast retrieval during inference.
    \item We introduce VeriFormer, which employs a set of learnable query tokens to extract code-relevant structural features from a graph database. Through a two-stage training strategy, VeriFormer aligns these graph embeddings with the LLM’s input embedding space, generating structure-aware soft prompts that improve the functional correctness of the generated Verilog code.
    \item We introduce VeriGRAG, a Verilog code generation framework that integrates graph retrieval with structure-aware soft prompting. With Qwen2.5-Coder-32B as its backbone, VeriGRAG achieves the best functional results among the compared methods on VerilogEval and RTLLM, measured by $pass@1$ and $pass@5$.
\end{itemize}



\section{Background and Motivation}

\subsection{LLM-Based Verilog Generation}

Recent work on LLM-based Verilog generation has focused on improving training data, generation reliability, and functional correctness. RTLCoder~\cite{liu2024rtlcoder} emphasizes dataset quality and expands its training corpus using GPT-generated examples. BetterV~\cite{pei2024betterv} constructs paired C--Verilog data from GitHub repositories and uses a generative discriminator to resample model outputs. AutoVCoder~\cite{gao2024autovcoder} iteratively generates hardware datasets and combines fine-tuning with retrieval-augmented generation. OriGen~\cite{cui2024origen} improves open-source LLMs through code-to-code augmentation and self-reflection, while MAGE~\cite{zhao2024mage} employs multiple agents to decompose and solve complex Verilog design tasks. HaVen~\cite{yang2025haven} targets HDL-specific hallucinations by aligning the generation framework with practical engineering workflows. More recent methods incorporate reasoning supervision and functional feedback. ReasoningV~\cite{qin2025reasoningv} introduces a functionally validated dataset and a two-stage training strategy that combines parameter-efficient and full-parameter tuning, whereas VeriPrefer~\cite{wang2025insights} uses simulator feedback and direct preference optimization to improve functional correctness.

\subsection{Motivation for Structure-Aware Retrieval}

Existing studies have explored text-based retrieval to enhance LLMs for Verilog generation. For example, AutoVCoder~\cite{gao2024autovcoder} retrieves supplementary hardware knowledge and incorporates it into the generation context. However, Verilog code represents circuits through explicit connectivity, hierarchy, and data dependencies, which are difficult to preserve in unstructured natural-language text. Moreover, when a target specification requires incremental modifications to an existing design, retrieved textual examples may dominate the generation process, potentially causing negative transfer or template copying. These limitations motivate a retrieval representation that exposes circuit structure rather than relying solely on natural language text. We propose VeriGRAG, which retrieves structural graphs from Verilog code and injects their aligned representations as soft prompts for structure-aware generation.



\section{Methodology}

In this section, we describe the key components of VeriGRAG. As illustrated in Figure~\ref{fig_overview}, the overall framework consists of three core components: (a) Encoding Verilog Graphs with GNN; (b) Knowledge-Distilled GraphEmb RAG; (c) Structure-Aware Soft Prompting with VeriFormer.

\subsection{Encoding Verilog Graphs with GNN}
We employ Yosys tools to extract data path graphs from Verilog code.
The Verilog code used in our experiments is sourced from the OriGen~\cite{cui2024origen} and PyraNet~\cite{nadimi2024pyranet} datasets, which form the basis of our graph database.
To ensure dataset quality, we use MinHash and Jaccard similarity to filter out duplicate or near-duplicate code samples.
For each Verilog module, we utilize the following Yosys command to extract its corresponding data path graph:
\begin{equation*}
    \texttt{read\_verilog;\allowbreak hierarchy -check;\allowbreak proc;\allowbreak opt;\allowbreak fsm}
\end{equation*}

\begin{figure}[tb!]
    \centering
    \includegraphics[width=1.0\columnwidth]{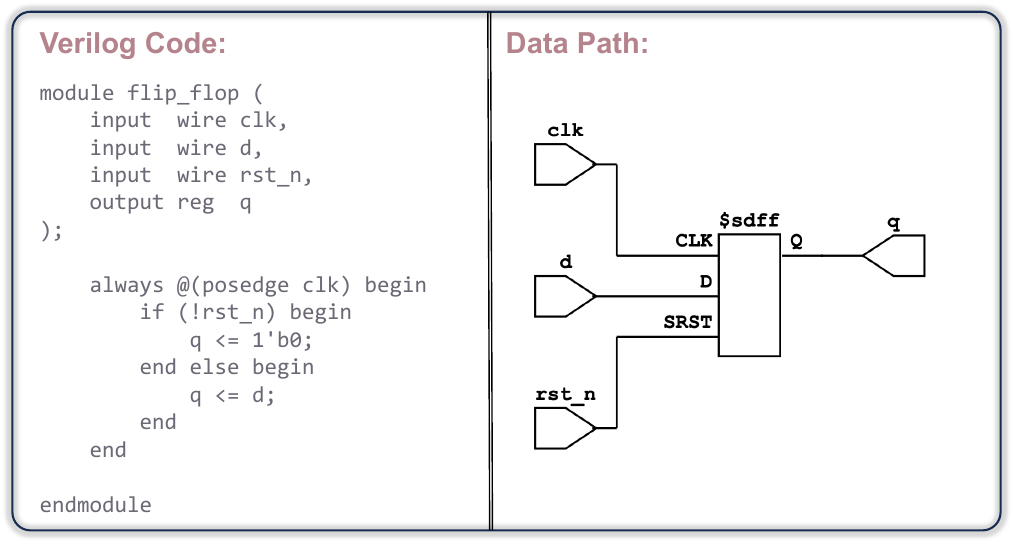}
    \caption{An example of data path graph from Verilog code. (a) Verilog code of a flip-flop; (b) the corresponding data path.}
    \Description{The left panel shows the Verilog code of a flip-flop, and the right panel shows its corresponding data path graph. Graph nodes represent hardware modules or signals, and edges represent their interconnections.}
    \label{fig_datapath}
\end{figure}

Each module is represented as a graph node, with inter-module connections explicitly defined by wires and modeled as edges. As illustrated in Figure~\ref{fig_datapath}, the left side shows the Verilog code of a flip-flop, while the right side depicts its corresponding data path graph. In the graph, nodes represent hardware modules or signals, and edges denote the interconnections between them. The node features include operation types, I/O types, I/O signal ports, and parameter lists, which are encoded into embedding representations using BERT~\cite{devlin2019bert}. Edge features are derived from normalized wire bit-widths and represent the structural relationships among connected modules. 

To extract graph features, we adopt the graph neural network of Graph Isomorphism Network with Edge features (GINEConv), which is modified from the Graph Isomorphism Network (GINConv)~\cite{Hu2020Strategies}. By integrating edge features that capture signal flows between hardware components, GINEConv effectively models data path dependencies in Verilog circuits and fully exploits the topological information of the graph structure, thereby extracting key features from data path more efficiently.
The GINEConv node aggregation function is defined as follows: 
\begin{equation}
    \mathbf{x}_i^{\prime} = h_{\Theta}\Big((1+\epsilon) \cdot \mathbf{x}_i + \sum_{j \in \mathcal{N}(i)} \mathrm{ReLU}(\mathbf{x}_j + \mathbf{e}_{j,i}) \Big).
\end{equation}

GINEConv effectively integrates edge weights into the message-passing mechanism during the aggregation process. Its parameters are initialized randomly and trained through an unsupervised contrastive learning framework. Specifically, given a graph $\widetilde{\mathcal{G}}$, two distinct augmented views, $\widetilde{\mathcal{G}}^{(1)}$ and $\widetilde{\mathcal{G}}^{(2)}$, are generated using stochastic data augmentation techniques such as edge deletion and feature perturbation. These two views are subsequently processed independently by GINEConv (denoted as $f_{\theta}$) to obtain their corresponding graph-level embedding representations.
\begin{equation}
    z_i^{(1)} = f_{\theta}(\widetilde{\mathcal{G}}^{(1)}),  z_i^{(2)} = f_{\theta}(\widetilde{\mathcal{G}}^{(2)}).
\end{equation}

We adopt the InfoNCE loss~\cite{oord2018representation} as our contrastive objective, which aims to maximize the similarity of positive pairs while minimizing it for negative samples. The loss is formulated as follows:
\begin{equation}
    Loss = 
    - \frac{1}{B}
    \sum_{i=1}^B 
    \log 
    \frac{
      \exp\left(\text{sim}\left( \mathbf{z_i^{(1)}}, \mathbf{z_i^{(2)}} \right) / \tau \right)
    }{
      \sum_{j=1}^B \exp\left( \text{sim}\left( \mathbf{z_i^{(1)}}, \mathbf{z_j^{(2)}} \right) /\tau \right)
    },
    \label{eq:infonce}
\end{equation}
where $B$ denotes batch size, $sim(\cdot, \cdot)$ is the cosine similarity function, and $\tau$ is the temperature scaling parameter that controls the concentration level of the softmax distribution. 

\subsection{Knowledge-Distilled GraphEmb RAG}

\begin{figure}[t]
    \centering
    \includegraphics[width=1.0\columnwidth]{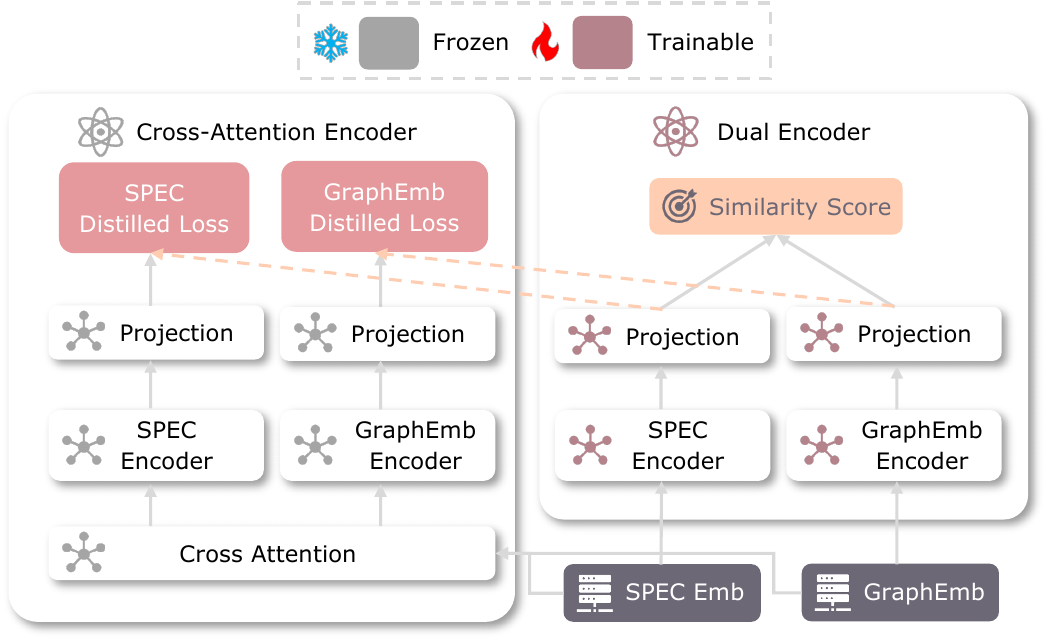}
    \caption{The knowledge distillation training process of our multimodal retriever}
    \Description{A cross-attention teacher encoder is first trained using the InfoNCE loss. With the teacher's parameters held fixed, a dual-encoder student is trained using InfoNCE and mean squared error losses to learn the teacher's embedding behavior. The student's independent encoders allow graph embeddings to be computed in advance and cached for efficient retrieval during inference.}
    \label{fig_retriever}
\end{figure}

To address the challenges of multimodal retrieval, we propose a cross-attention encoder architecture, which adds a shared cross-attention layer before the conventional dual encoder framework, enabling bidirectional interaction between the two modalities. Preliminary experiments suggest that this cross-attention encoder effectively enhances the similarity between positive samples during training. However, due to the presence of the shared cross-attention layer, pre-encoding of graph embeddings becomes infeasible, making it impossible to cache graph embeddings. As a result, retrieval efficiency inevitably degrades when operating on large-scale graph database.

To balance retrieval accuracy with efficiency, we introduce a knowledge distillation framework. As illustrated in Figure~\ref{fig_retriever}, we first train a cross-attention encoder (teacher model) using unsupervised contrastive learning with the InfoNCE loss (Equation~\ref{eq:infonce}). After freezing the teacher model's parameters, we train a dual encoder (student model) to mimic the teacher model's embedding behavior. The student model is optimized using a combination of (1) the InfoNCE loss constructed from positive/negative samples, and (2) a mean squared error (MSE) loss relative to the teacher model’s output. This knowledge distillation training strategy preserves the retrieval efficiency of the dual encoder architecture while approximating the accuracy of the cross-attention encoder. 

At inference time, the independence of the two encoders enables pre-encoding and indexing of graph embeddings using FAISS~\cite{johnson2019billion}, a highly efficient similarity search library for large-scale vector database. User queries are encoded in real-time and efficiently retrieved against the cached graph embeddings, enabling fast and scalable retrieval over large-scale graph databases.

\subsection{Structure-Aware Soft Prompting with VeriFormer}

\begin{figure}[t]
    \centering
    \includegraphics[width=1.0\columnwidth]{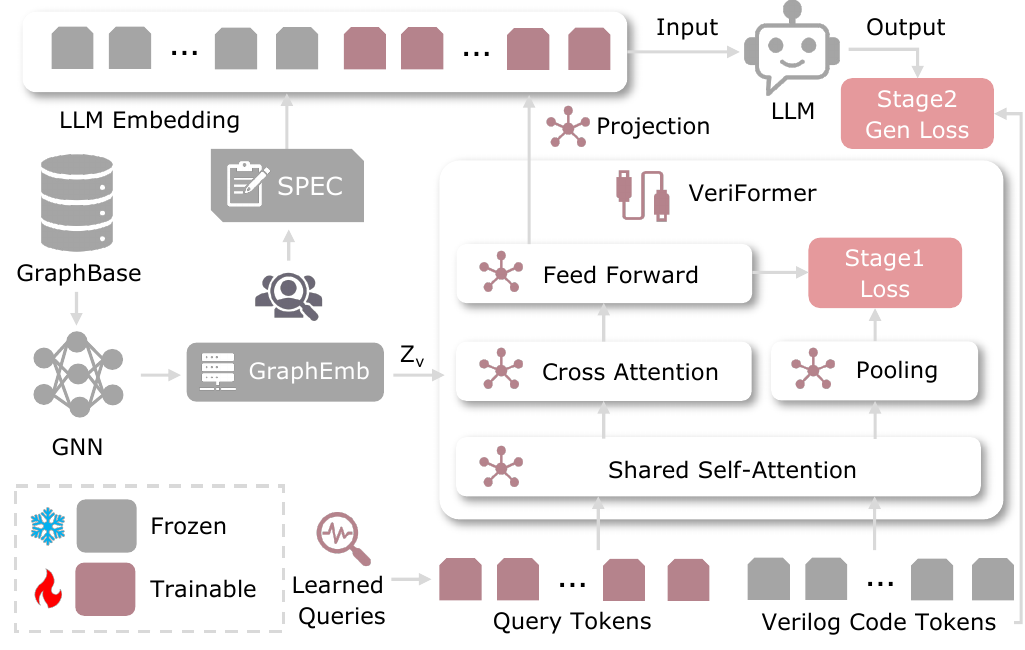}
    \caption{
        The overall framework of VeriFormer aligns the GNN and the LLM through a two-stage training paradigm to optimize Verilog code generation.
        The lightweight VeriFormer module is trained using this two-stage paradigm. 
    }
    \label{fig_veriformer}
    \Description{VeriFormer is trained in two stages. In the first stage, the GNN is frozen, while VeriFormer and learnable query tokens are trained with graph-code contrastive learning, generation, and matching objectives. In the second stage, the code transformer is discarded, and graph-transformer query outputs are projected into the LLM's input embedding space as soft prompts. Training combines the LLM generation loss with a KL-divergence loss for distribution alignment.}
\end{figure}

Inspired by~\cite{li2023blip}, we introduce the VeriFormer module to align the input text modality of the LLM with the retrieved graph embedding modality, as illustrated in Figure~\ref{fig_veriformer}.

The VeriFormer module consists of two transformer sub-modules that share the same self-attention layers. We introduce a graph transformer and a code transformer within VeriFormer to align graph embeddings with the input text embeddings of the LLM. In this framework, the graph transformer interacts with the frozen GNN, while the code transformer processes Verilog code. The graph transformer takes a set of learnable query tokens as input, which interact through the shared self-attention layers and are aligned with the graph embeddings generated by the frozen GNN through cross-attention layers. Additionally, these query tokens also interact with the code transformer through the shared self-attention layers. We initialize VeriFormer using pre-trained weights from CodeBERT~\cite{feng2020codebert}. In our experiments, we employ 96 query tokens, each with a dimension of 768, matching the hidden layer dimension of VeriFormer.

We employ a two-stage training paradigm to train the lightweight VeriFormer model.

In the first training stage, we train VeriFormer to extract graph embeddings that are most relevant to the Verilog code from the outputs of the GNN. During this stage, only VeriFormer and the learnable query tokens are trainable, while the GNN parameters are kept frozen. 
We optimize three objectives to align the graph embeddings with the Verilog code representations: 
\textbf{(1) Graph–Code Contrastive Learning Objective.}
This objective aligns the query embeddings $\mathcal{G}$ from the graph transformer with the pooled code representation $\mathcal{C}$ from the code transformer. It maximizes the similarity between positive graph–code pairs while contrasting them against negatives to encourage discriminative alignment.
\textbf{(2) Graph–Code Generation Objective.}
This objective trains VeriFormer to generate Verilog code conditioned on graph embeddings through learnable query tokens and cross-attention. The model is optimized with cross-entropy loss to encourage the query tokens to capture code-relevant structural semantics.
\textbf{(3) Graph–Code Matching Objective.}
This objective performs fine-grained binary classification to predict whether a graph–code pair matches. For each pair, query-level logits are aggregated to compute the final matching score, enabling instance-level alignment between the modalities.

In the second training stage, we keep only the graph transformer module and discard the code transformer module.
The learnable query tokens are further trained, and the query outputs are projected through a projection layer to match the embedding dimension of the LLM input.
This projected output is then used as a soft prompt for the LLM.

Because the projected soft prompts may differ in distribution from the LLM token embeddings, we introduce a KL-divergence loss to align them. For distribution alignment, we randomly sample LLM input embeddings without replacement to match the number of soft-prompt tokens. The specific computation is as follows:
\begin{equation}
    Loss_{Dist} = \frac{1}{B\cdot N} \sum^{B\cdot N}_{i=1}\sum^{D}_{j=1} \mathbf{Z_i^{(j)}}
    \left(\log \mathbf{Z_i^{(j)}} - \log \mathbf{G_i^{(j)}} \right),
    \label{eq:kl_loss}
\end{equation}
where $\mathbf{Z}$ and $\mathbf{G}$ denote the softmax-normalized distributions of the sampled, detached LLM input embeddings and the projected VeriFormer soft prompts, respectively. $B$ denotes the batch size, $N$ denotes the number of soft-prompt tokens, and $D$ denotes the LLM input embedding dimension.

Thus, the loss function in the second stage consists of two components: the generation loss of the LLM and the distribution loss, as formally computed below:
\begin{equation}
    Loss = Loss_{Gen} + \alpha \cdot Loss_{Dist},
\end{equation}
where $\alpha$ denotes the weight coefficient for the distribution loss, which balances its contribution relative to the generation loss during training.



\begin{table*}[!t]
  \centering
  \footnotesize
  \caption{Comparison of Model Performance On VerilogEval-Machine, VerilogEval-Human and VerilogEval v2.}
 
    \begin{tabular}{|ccc|cc|cc|cccc|}
        \hline
        \multirow{3}[0]{*}{{Type}} & \multirow{3}[0]{*}{{Model}} & \multirow{3}[0]{*}{{Params}} & 
        \multicolumn{2}{c|}{{VerilogEval-Machine}} & \multicolumn{2}{c|}{{VerilogEval-Human}} & \multicolumn{4}{c|}{{VerilogEval v2}} \\
        &   &   & \multicolumn{2}{c|}{{Function}} & \multicolumn{2}{c|}{{Function}} & \multicolumn{2}{c}{{Syntax}} & \multicolumn{2}{c|}{{Function}} \\
        \cline{4-11}
      &   &   & $pass@1$ & $pass@5$ & $pass@1$ & $pass@5$ & $pass@1$ & $pass@5$ & $pass@1$ & $pass@5$ \\
      \hline \hline
      \multirow{4}[0]{*}{{General LLMs}} & GPT-4o~\cite{hurst2024gpt} & N/A & 65.9 & 71.4 & 57.1 & 63.9 & 90.7 & 95.2 & 56.5 & 65.1 \\
      & Llama3.1~\cite{grattafiori2024llama}  & 405B & 67.3 & 75.1 & 53.8 & 61.0 & 91.2 & 93.5 & 50.2 & 57.1 \\
      & CodeLlama~\cite{roziere2023code}  & 7B & 43.1 & 47.1 & 18.2 & 22.7 & 37.1 & 56.1 & 12.1 & 20.2 \\
      & CodeQwen~\cite{qwen}  & 7B & 46.5 & 54.9 & 22.5 & 26.1 & 52.4 & 57.0 & 19.8 & 24.3 \\
      \hline
      \multirow{2}[0]{*}{{RTLCoder~\cite{liu2024rtlcoder}}} & Mistral & 7B & 62.5 & 72.2 & 36.7 & 45.5 & 69.8 & 78.1 & 34.0 & 37.5 \\
      & DeepSeek-Coder  & 6.7B & 61.2 & 76.5 & 41.6 & 50.1 & 87.0 & 94.6 & 40.9 & 47.9 \\
      \hline
      \multirow{3}[0]{*}{{BetterV~\cite{pei2024betterv}}} & CodeLlama & 7B & 64.2 & 75.4 & 40.9 & 50.0 & N/A & N/A & N/A & N/A \\
      & DeepSeek-Coder & 6.7B & 67.8 & 79.1 & 45.9 & 53.3 & N/A & N/A & N/A & N/A \\
      & CodeQwen & 7B & 68.1 & 79.4 & 46.1 & 53.7 & N/A & N/A & N/A & N/A \\
      \hline
      \multirow{3}[0]{*}{{AutoVCoder~\cite{gao2024autovcoder}}} & CodeLlama & 7B & 63.7 & 72.9 & 44.5 & 52.8 & N/A & N/A & N/A & N/A \\
      & DeepSeek-Coder & 6.7B & 69.0 & 79.3 & 46.9 & 53.7 & N/A & N/A & N/A & N/A \\
      & CodeQwen & 7B & 68.7 & 79.9 & 48.5 & 55.9 & N/A & N/A & N/A & N/A \\
      \hline
      \multirow{3}[0]{*}{{CodeV~\cite{zhao2024codev}}} & CodeLlama & 7B & 78.1 & 86.0 & 45.2 & 59.5 & 93.7 & 95.2 & 46.1 & 52.5 \\
      & DeepSeek-Coder & 6.7B & 77.9 & \underline{88.6} & 52.7 & 62.5 & 93.3 & 96.8 & 47.1 & 54.4 \\
      & CodeQwen & 7B & 77.6 & 88.2 & 53.2 & 65.1 & 95.4 & 95.9 & 48.4 & 56.1 \\
      \hline
      {OriGen~\cite{cui2024origen}} & DeepSeek-Coder & 6.7B & 74.1 & 82.4 & 43.3 & 46.4 & 77.5 & 87.8 & 51.3 & 56.3 \\
      \hline
      \multirow{3}[0]{*}{{HaVen~\cite{yang2025haven}}} & CodeLlama & 7B & 74.7 & 80.0 & 47.5 & 54.6 & 88.2 & 94.8 & 41.8 & 47.9 \\
      & DeepSeek-Coder & 6.7B & 78.8 & 84.5 & 46.6 & 56.6 & 97.1 & 98.9 & 47.2 & 52.3 \\
      & CodeQwen & 7B & 77.3 & 81.2 & 53.3 & 57.8 & 97.0 & 98.5 & 47.6 & 51.9 \\
      \hline
      \multirow{3}[0]{*}{{VeriPrefer~\cite{wang2025insights}}} & Qwen2.5-Coder & 7B & 72.7 & 85.8 & 49.7 & 62.3 & 95.6 & \underline{99.9} & 49.3 & 64.7 \\
      & DeepSeek-Coder & 6.7B & 71.2 & 81.5 & 52.8 & 64.0 & 98.3 & 99.8 & 52.1 & 58.5 \\
      & Qwen2.5-Coder & 14B & \underline{81.8} & 87.3 & \underline{61.9} & \underline{71.3} & 96.8 & \underline{99.9} & 58.6 & \underline{72.1} \\
      \hline
      \multirow{3}[0]{*}{{{VeriGRAG (Ours)}}} & Qwen2.5-Coder & 7B & 79.4 & 87.3 & 56.4 & 65.6 & \underline{99.5} & \textbf{100.0} & \underline{60.4} & 66.8\\
      & DeepSeek-Coder & 7B & 78.0 & 86.1 & 51.1 & 61.0 & \underline{99.5} & \textbf{100.0} & 52.3 & 62.0 \\
      & Qwen2.5-Coder & 32B & \textbf{85.1} & \textbf{90.2} & \textbf{62.9} & \textbf{72.0} & \textbf{99.9} & \textbf{100.0} & \textbf{66.8} & \textbf{75.2} \\
      \hline
    \end{tabular}
    
  \label{tab:veri_eval}
\end{table*}

\section{Evaluation}
\subsection{Experimental Setup}
For pre-trained models, we selected DeepSeek-Coder-7B-Instruct-v1.5~\cite{guo2024deepseek}, as well as Qwen2.5-Coder-7B-Instruct and Qwen2.5-Coder-32B-Instruct~\cite{hui2024qwen2}.

We employ low-rank adaptation (LoRA)~\cite{hu2022lora} to train the model's capability for generating RTL code.
The OriGen~\cite{cui2024origen} and PyraNet~\cite{nadimi2024pyranet} Verilog code datasets are selected as training sources due to their complementary coverage of diverse hardware description tasks,
which enables the LLM to learn structural patterns from a wide range of circuit designs.
MinHash and Jaccard similarity are utilized to eliminate duplicate samples, followed by additional filtering to remove entries with empty or meaningless descriptions and code segments.
A final dataset comprising 276,627 examples is obtained after the preprocessing steps.

Fine-tuning is performed on a machine with eight NVIDIA H100-80GB HBM3 graphics cards with CUDA driver 12.8 for 4 epochs.
For all the models, we use the AdamW optimizer~\cite{loshchilov2018decoupled} with parameters $\beta_1=0.9$ and $\beta_2=0.999$, along with cosine learning rate decay for scheduling.
The warm-up ratio is set to 0.03 and weight decay is set to 0.1 with a learning rate of 5e-5.
The global batch size is set to 128.

As for RAG retriever training, the teacher model cross-attention encoder is trained for 15 epochs at a learning rate of 5e-5, while the student model dual encoder is trained for 100 epochs at a learning rate of 5e-4 and the global batch size is set to 1024.
In the first training stage of VeriFormer, the model is trained for 3 epochs at a learning rate of 1e-4 with a minimum learning rate of 1e-6 and the global batch size is set to 400.
For the second training stage of VeriFormer, the model is trained with frozen LLM weights for 1 epoch at a learning rate of 5e-6, with an early stopping strategy applied and the global batch size set to 64.

Following HAVEN~\cite{yang2025haven} and VeriPrefer~\cite{wang2025insights}, we reported optimal results in three temperature settings: 0.2, 0.5, and 0.8.

\subsection{Benchmark and Evaluation Metrics}
To evaluate the performance of Verilog code generation, we select two representative benchmarks: VerilogEval v1~\cite{liu2023verilogeval} and RTLLM v1.1~\cite{lu2024rtllm}. VerilogEval is derived from approximately 150 Verilog tasks on the HDLBits website, which were manually converted to create VerilogEval-Human, with VerilogEval-Machine generated by GPT-3.5-Turbo~\cite{openai2023gpt35}. RTLLM v1.1 consists of 29 Verilog tasks with a wider range of difficulty levels, closely related to real-world design tasks. In addition, we also conducted an evaluation of the newly released versions of these two benchmarks: VerilogEval v2~\cite{pinckney2024revisiting} and RTLLM v2~\cite{liu2024openllm}. VerilogEval v2 builds on the original VerilogEval-Human and extends its scope to encompass specification-to-RTL design tasks, with a prompt style resembling that of a chatbot. RTLLM v2 expands RTLLM v1.1 to 50 designs in four categories: Arithmetic, Memory, Control, and Miscellaneous.

Both benchmarks use the widely adopted $pass@k$ metric to evaluate the functional and syntactic correctness of the generated code. In this metric, if any of the $k$ samples pass the corresponding check, the problem is considered solved. It is expressed as:
\begin{equation}
    \textit{pass@k} := \mathbb{E}\left[ 1 - \frac{{n-c \choose k}}{{n \choose k}} \right],
\end{equation}
where $n \geq k$ denotes the total number of trials per problem, and $c$ represents the number of trials that passed the syntax or functional check. We set $n=20$ in experiments and report results for $k=1$ and $k=5$.

\subsection{Comparison with Existing Methods}
Baseline results in Tables~\ref{tab:veri_eval} and~\ref{tab:rtllm} are primarily taken from VeriPrefer~\cite{wang2025insights}, which re-evaluated open-source methods. In these two tables, bold and underlined values indicate the best and second-best results, respectively. Among the DeepSeek-Coder-based methods, VeriGRAG uses the v1.5 (7B) backbone, while the baselines are listed with v1 (6.7B). The ablation studies in Tables~\ref{tab:verilogeval_v1} and~\ref{tab:verilogeval_v2} provide controlled comparisons using the same backbone and retrieval database.

Tables~\ref{tab:veri_eval} and~\ref{tab:rtllm} show that VeriGRAG consistently matches or exceeds competitive open-source and proprietary models in both syntactic and functional correctness. The advantage is particularly evident on VerilogEval-Human and VerilogEval v2, whose tasks and prompts better reflect practical HDL design requirements. On VerilogEval v2, VeriGRAG with the Qwen2.5-Coder-32B backbone achieves the highest functional correctness, with $pass@1$ and $pass@5$ scores of $66.8\%$ and $75.2\%$, respectively. Among the competing methods using the same Qwen2.5-Coder-7B backbone, VeriGRAG also achieves the best performance, reaching $60.4\%$ in $pass@1$ and $66.8\%$ in $pass@5$.

On RTLLM v2, VeriGRAG achieves syntactic $pass@1$ scores of $98.0\%$--$98.3\%$ across the three evaluated backbones, compared with $90.6\%$ for the strongest baseline on this metric. For functional correctness, VeriGRAG with Qwen2.5-Coder-32B reaches $66.1\%$ in $pass@1$, the highest among the compared methods. With the same Qwen2.5-Coder-7B backbone, VeriGRAG achieves a functional $pass@1$ of $56.7\%$, exceeding VeriPrefer's $52.4\%$ by 4.3 percentage points.

These results support the effectiveness of incorporating retrieved structural information through structure-aware soft prompts to improve both syntactic and functional correctness in Verilog generation.

\subsection{Ablation Study}

\begin{table*}[tb!]
  \centering
  \footnotesize
  \caption{Comparison of Model Performance On RTLLM v1.1 and RTLLM v2.}
  \label{tab:rtllm}
  
      \begin{tabular}{|ccc|cc|cc|cccc|}
          \hline
          \multirow{3}[0]{*}{{Type}} & \multicolumn{1}{c}{\multirow{3}[0]{*}{{Model}}} & \multicolumn{1}{c|}{\multirow{3}[0]{*}{{Params}}} & \multicolumn{4}{c|}{{RTLLM v1.1}} & \multicolumn{4}{c|}{{RTLLM v2}} \\
          &   &   & \multicolumn{2}{c|}{{Syntax}} & \multicolumn{2}{c|}{{Function}} & \multicolumn{2}{c}{{Syntax}} & \multicolumn{2}{c|}{{Function}} \\
          \cline{4-11}
          &   &   & $pass@1$ & $pass@5$ & $pass@1$ & $pass@5$ & $pass@1$ & $pass@5$ & $pass@1$ & $pass@5$ \\
          \hline \hline
          \multirow{4}[0]{*}{{General LLMs}} & {GPT-4o~\cite{hurst2024gpt}} & {N/A} & 82.4 & 86.2 & 47.9 & 58.0 & 80.0 & 89.5 & 47.9 & 58.0 \\
          & {Llama3.1~\cite{grattafiori2024llama}} & {405B} & 73.2 & 81.8 & 38.9 & 45.8 & 73.0 & 81.6 & 38.8 & 45.6 \\
          & CodeLlama~\cite{roziere2023code}  & 7B & 46.6 & 62.6 & 17.9 & 29.9 & 49.5 & 76.7 & 21.2 & 31.9 \\
          & CodeQwen~\cite{qwen}  & 7B & 45.8 & 65.8 & 24.1 & 34.0 & 47.1 & 66.4 & 25.8 & 29.0 \\
          \hline
          \multirow{2}[0]{*}{{RTLCoder~\cite{liu2024rtlcoder}}} & {Mistral} & {7B} & 64.6 & 73.7 & 24.5 & 37.3 & 75.6 & 81.1 & 37.0 & 39.9 \\
          & DeepSeek-Coder & {6.7B} & 73.4 & 83.9 & 35.8 & 40.3 & 82.9 & 90.0 & 43.5 & 48.0 \\
          \hline
          \multirow{3}[0]{*}{{CodeV~\cite{zhao2024codev}}} & {CodeLlama} & {7B} & 79.0 & 89.2 & 39.4 & 50.3 & 76.8 & 91.3 & 47.4 & 53.3 \\
          & DeepSeek-Coder & {6.7B} & 78.3 & 87.4 & 42.4 & 51.5 & 75.5 & 90.7 & 45.5 & 55.7 \\
          & CodeQwen & {7B} & 78.8 & 89.5 & 36.6 & 53.3 & 78.1 & 89.6 & 48.1 & 56.9 \\
          \hline
          {OriGen~\cite{cui2024origen}} & {DeepSeek-Coder} & {6.7B} & 78.1 & 86.4 & 45.2 & 58.4 & 77.5 & 87.8 & 40.9 & 57.1 \\
          \hline
          \multirow{3}[0]{*}{{HaVen~\cite{yang2025haven}}} & {CodeLlama} & {7B} & 45.5 & 95.4 & 42.3 & 46.8 & 51.7 & 92.4 & 44.9 & 48.2 \\
          & {DeepSeek-Coder} & {6.7B} & 88.9 & 92.8 & 45.4 & 55.3 & 89.0 & 94.2 & 50.6 & 60.5 \\
          & {CodeQwen} & {7B} & 87.6 & 92.8 & 45.1 & 53.3 & 88.6 & 94.9 & 51.2 & 59.7 \\
          \hline
          \multirow{3}[0]{*}{{VeriPrefer~\cite{wang2025insights}}} &  Qwen2.5-Coder & 7B & 90.1 & 93.8 & 53.2 & 67.7 & 90.1 & 99.6 & 52.4 & 66.4 \\
          & DeepSeek-Coder & 6.7B & 87.4 & 96.1 & 50.7 & 64.7 & 90.6 & 97.6 & 50.4 & 63.0 \\
          & Qwen2.5-Coder & 14B & 90.5 & 95.2 & 50.2 & 63.9 & 85.2 & \underline{99.7} & 55.2 & \underline{67.5} \\
          \hline
          \multirow{3}[0]{*}{{{VeriGRAG (Ours)}}} & {Qwen2.5-Coder} & 7B & \underline{93.8} & \textbf{96.6} & 57.2 & \underline{69.3} & 98.0 & \textbf{100.0} & 56.7 & 66.5 \\
          & {DeepSeek-Coder} & 7B & \textbf{94.7} & \textbf{96.6} & \underline{62.2} & 68.5 & \underline{98.2} & \textbf{100.0} & \underline{61.2} & 67.2 \\
          & Qwen2.5-Coder & 32B & \underline{93.8} & \underline{96.5} & \textbf{62.8} & \textbf{75.4} & \textbf{98.3} & \textbf{100.0} & \textbf{66.1} & \textbf{74.5} \\
          \hline
      \end{tabular}

\end{table*}

\begin{figure*}[th]
    \centering
    \subfloat[VerilogEval v1 Machine and Human test]{%
        \includegraphics[width=0.49\linewidth]{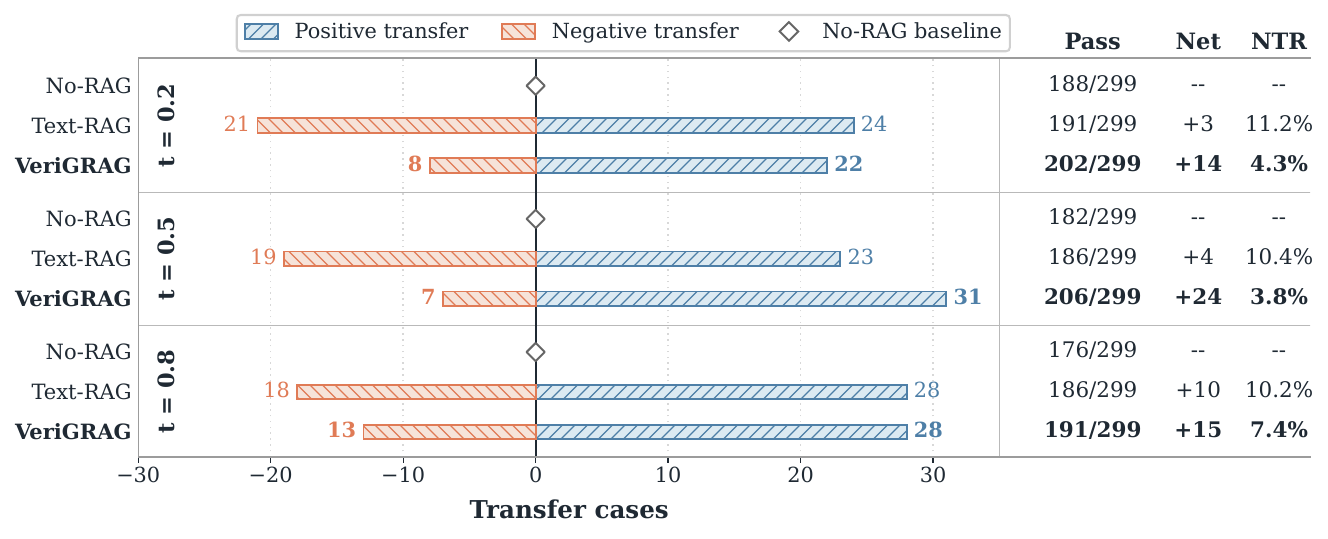}
        \label{fig:negative_transfer-a}
    }
    \hfill
    \subfloat[Controlled modification stress test]{%
        \includegraphics[width=0.49\linewidth]{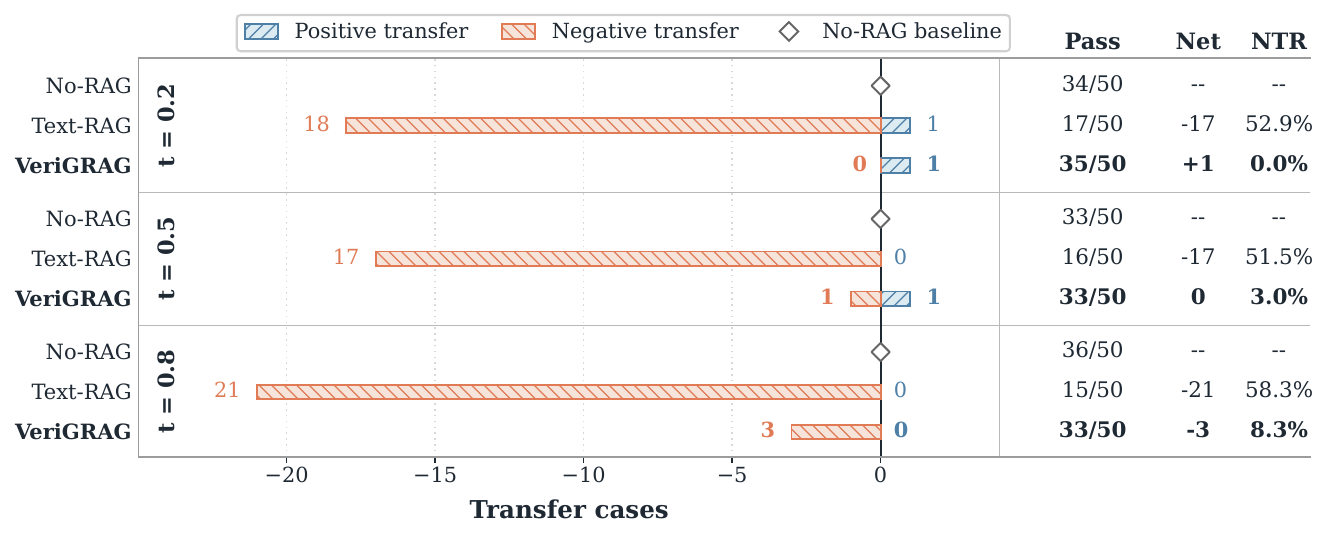}
        \label{fig:negative_transfer-b}
    }
    \caption{Positive and negative transfers at different temperatures, measured relative to No-RAG. NTR denotes the negative transfer rate, defined as the proportion of No-RAG passing cases that fail after retrieval.}
    \Description{Two panels compare retrieval-induced transfer for VeriGRAG and Text-RAG at temperatures 0.2, 0.5, and 0.8 using Qwen2.5-Coder-7B. Panel (a) evaluates 299 VerilogEval-Machine and VerilogEval-Human samples: VeriGRAG yields larger net gains and fewer negative transfers at all three temperatures. Panel (b) evaluates 50 RTL designs with localized modifications and conflicting retrieval context: VeriGRAG reduces negative transfers at all three temperatures. Positive transfer corrects a No-RAG failure; negative transfer causes a No-RAG passing case to fail.}
    \label{fig:negative_transfer}
\end{figure*}

Tables~\ref{tab:verilogeval_v1} and~\ref{tab:verilogeval_v2} assess structure-aware soft prompting and knowledge distillation (KD) with Qwen2.5-Coder-7B and the same retrieval database. Text-RAG uses retrieved raw text as the prompt, whereas VeriGRAG uses structure-aware soft prompts.

Without KD, VeriGRAG improves $pass@1$ over Text-RAG from $53.8\%$ to $56.0\%$ on VerilogEval-Human and from $58.2\%$ to $60.4\%$ on VerilogEval v2. Adding KD further improves $pass@1$ from $78.4\%$ to $79.4\%$ and $pass@5$ from $86.3\%$ to $87.3\%$ on VerilogEval-Machine. On VerilogEval-Human, the corresponding scores increase from $56.0\%$ to $56.4\%$ and from $63.8\%$ to $65.6\%$. On VerilogEval v2, KD increases $pass@5$ from $65.5\%$ to $66.8\%$, while $pass@1$ remains at $60.4\%$.

Overall, VeriGRAG improves functional correctness over Text-RAG by using structure-aware soft prompts. Distilling the retriever yields further gains in most evaluated settings, suggesting that KD improves the relevance of the retrieved structural information for generating functionally correct Verilog code.

\begin{table}[t]
    \centering
    \footnotesize
    \caption{Functional correctness results on VerilogEval v1.}
    \label{tab:verilogeval_v1}
    {
    \begin{tabular}{|c|cc|cc|}
        \hline
        \multirow{2}{*}{Method}
        & \multicolumn{2}{c|}{Machine}
        & \multicolumn{2}{c|}{Human} \\
        & $pass@1$ & $pass@5$ & $pass@1$ & $pass@5$ \\
        \hline \hline
        Text-RAG
        & 77.2 & 86.6 & 53.8 & 62.5 \\
        VeriGRAG w/o KD
        & 78.4 & 86.3 & 56.0 & 63.8 \\
        VeriGRAG w/ KD
        & \textbf{79.4} & \textbf{87.3}
        & \textbf{56.4} & \textbf{65.6} \\
        \hline
    \end{tabular}
    }
\end{table}

\begin{table}[t]
    \centering
    \footnotesize
    \caption{Functional correctness results on VerilogEval v2.}
    \label{tab:verilogeval_v2}
    {
    \begin{tabular}{|c|cc|}
        \hline
        Method & $pass@1$ & $pass@5$ \\
        \hline \hline
        Text-RAG
        & 58.2 & 65.3 \\
        VeriGRAG w/o KD
        & \textbf{60.4} & 65.5 \\
        VeriGRAG w/ KD
        & \textbf{60.4} & \textbf{66.8} \\
        \hline
    \end{tabular}
    }
\end{table}

\subsection{Robustness to Negative Transfer}

We evaluate retrieval-induced transfer effects using Qwen2.5-Coder-7B at temperatures $t\in \{0.2,0.5,0.8\}$. For this analysis, each case is evaluated using a single generation, with pass/fail determined by its testbench. The evaluation covers the combined 299 samples from VerilogEval-Machine (143 samples) and VerilogEval-Human (156 samples)~\cite{liu2023verilogeval}. Positive transfer occurs when retrieval corrects a sample that fails under No-RAG, whereas negative transfer occurs when retrieval causes a previously passing sample to fail.
As shown in Figure~\ref{fig:negative_transfer-a}, VeriGRAG consistently reduces negative transfers and achieves larger net gains than Text-RAG across all three temperatures. This result indicates that VeriGRAG is less sensitive to misleading retrieval information and more robust to retrieval-induced negative transfer.

To further examine this robustness, we construct a controlled stress test comprising 50 RTL designs with localized incremental modifications.
Each sample changes one localized behavior while preserving the remaining functionality, and the original implementation is provided as conflicting retrieval context.
As shown in Figure~\ref{fig:negative_transfer-b}, Text-RAG frequently retains obsolete implementation details,
whereas VeriGRAG substantially reduces negative transfers across all temperatures.
By focusing on retrieved structural information rather than the original design, VeriGRAG is less sensitive to misleading retrieval and negative transfer.



\section{Conclusion}
This paper presents VeriGRAG, a structure-aware soft prompt framework for Verilog code generation. Data path graphs extracted by Yosys are encoded with a GNN and stored for retrieval. A dual encoder, trained via knowledge distillation from a cross-attention encoder, retrieves relevant graph embeddings for a given hardware description. VeriFormer then uses learnable query tokens and two-stage training to extract code-relevant structural features and align them with the LLM's input embedding space, producing structure-aware soft prompts. Experiments on VerilogEval and RTLLM show improvements in syntactic and functional correctness. With Qwen2.5-Coder-32B, VeriGRAG achieves the highest functional $pass@1$ and $pass@5$ scores among the compared methods on both benchmarks.


\begin{acks}
This work was supported in part by the Strategic Priority Research Program of the Chinese Academy of Sciences (CAS) under Grants XDB0660101, XDB0660100, and XDB0660000, and in part by the National Natural Science Foundation of China (NSFC) under Grants 92473114 and 62672470.
\end{acks}

\bibliographystyle{ACM-Reference-Format}




\end{document}